# Disrupting Terrorist Networks – A Dynamic Fitness Landscape Approach

Philip V. Fellman, School of Business
Southern New Hampshire University
Shirogitsune99@yahoo.com

Jonathan P. Clemens,
Intel Corporation
Jonathan.p.clemens@intel.com

Roxana Wright
Keene State College
Rox_wright@yahoo.com

Jonathan Vos Post,
Computer Futures, Inc.
Jvospost2@yahoo.com

Matthew Dadmun
Southern New Hampshire University
mangell68@hotmail.com

## Introduction

The study of terrorist networks as well as the study of how to impede their successful functioning has been the topic of considerable attention since the odious event of the 2001 World Trade Center disaster. While serious students of terrorism were indeed engaged in the subject prior to this time, a far more general concern has arisen subsequently. Nonetheless, much of the subject remains shrouded in obscurity, not the least because of difficulties with language and the representation or translation of names, and the inherent complexity and ambiguity of the subject matter.

One of the most fruitful scientific approaches to the study of terrorism has been network analysis (Krebs, 2002; Carley, 2002a; Carley and Dombroski, 2002; Butts, 2003a; Sageman, 2004, etc.) As has been argued elsewhere, this approach may be particularly useful, when properly applied, for disrupting the flow of communications ($C^4I$) between levels of terrorist organizations (Carley, Krackhardt and Lee, 2001; Carley, 2002b; Fellman and Wright, 2003; Fellman and Strathern, 2004; Carley et al, 2003; 2004). In the present paper we examine a recent paper by Ghemawat and Levinthal, (2000) applying Stuart Kauffman's NK-Boolean fitness landscape approach to the formal mechanics of decision theory. Using their generalized NK-simulation approach, we suggest some ways in which optimal decision-making for terrorist networks might be constrained and following our earlier analysis, suggest ways in which the forced compartmentation of terrorist organizations by counter-terrorist security organizations might be more likely to impact the quality of terrorist organizations' decision-making and command execution.

## General Properties of Terrorist Networks

Without attempting to be either exhaustive or exhausting, recent research on terrorism has revealed several relevant characteristics of terrorist organizations which a prudent modeler ought to keep in mind. These networks are first and foremost, *covert,* which means that they have hidden properties, and our information about them is necessarily incomplete, hence demanding complex methodological tools for determining the properties of the network structure (Butts, 2001, 2003a; Carley 2002a, 2003; Krebs, 2001, Clemens and O'Neill, 2004). While we are primarily concerned in the present paper with formal properties of terrorist networks, it does bear keeping in mind that at the operational level they are *purposive*, which lends them not only formal characteristics, but depending upon the organization in question, a considerable ideological history (Hoffman, 1997; Hoffman and Carr, 1997), and in some cases, rather serious (path-dependent) constraints on recruiting (Codevilla, 2004a; Fellman and Strathern, 2004) targets, and methods (Sageman, 2004). Some other, rather interesting properties of terrorist networks include the fact that they are often separated by larger than normal degrees of distance between their participants, a condition arising from their covert nature (Krebs, 2001; Fellman and Wright, 2003; Carley, 2003). Curiously, this kind of structure appears to have an emergent shape, which can be mapped as a distributed network (Krebs, 2001; Fellman and Strathern, 2004; Clemens and O'Neill, 2004), commonly illustrated by a social network diagram of the 9-11 Hijackers and informally referred to as "the dragon".

Carley et al (2001,2002b, 2003, 2004) have developed useful models for distinguishing cohesive vs. adhesive organizations as well as defining probably outcomes for the removal of higher visibility nodes. Formal models of network analysis can also suggest where removal of key nodes or vertices can disrupt the organization's ability to transmit commands across hierarchical levels of the organization, thus leading to command degeneration (Butts, 2003a; Carley et al, 2004). The difficulty with this approach is that an important aspect of the dynamics of terrorist networks is that they are learning organizations (Hoffman, 1997; Tsvetovat and Carley, 2003).

If one bears all of these features in mind, some of the complexities of dealing with terrorist organizations become immediately apparent. Terrorists are slippery foes, they are hidden, they have redundant command structures, they change their membership (not all of which changes are visible) and they learn from their mistakes. Nobody who has to deal with terrorist threats wants to see those threats and the organizations that make them, evolve. The obviousness of this proposition is evidenced by the U.S. reaction to 9-11. What then, are the possible approaches?

Complexity science has afforded a number of approaches to evolution in general (Kauffman, 1993, 1996, 2000) as well as to the evolution of organizations and the ways in which complexity science may be applied to problems of organizational behavior. In particular, Kauffman's NK-Boolean fitness landscape model appears to offer a number of fruitful heuristics (Lissack, 1996; McKelvey, 1999; Meyer, 1996; Fellman et al, 2004). In 1999, seeking to define the formal properties of an optimal business organization decision-making process, Pankaj Ghemawat of Harvard Business School and Daniel Levinthal of the Wharton School ran an agent based simulation of decision-making in order to define the ways in which decisional interdependence and the interdependence of business units affect overall performance (fitness). In the section which follows, we will explore a number of their findings and suggest how they might be applied to inhibiting the fitness of terrorist organizations.

## The Structure of the Ghemawat-Levinthal NK Simulation

A primary goal of the simulation was to model interdependent choices. Levinthal and Ghemawat focus on this aspect of decision making because they are attempting to understand the formal structure of decision-making in organizations with interdependent parts. While they rapidly come to focus on the same measures that we have seen used to characterize terrorist networks, hierarchy and centrality (Butts, 2001, 2003; Carley et al, 2001; 2003, 2004; Carley, 2002, 2003), plus an additional factor of randomness (which most of us are wont to deny) they come at these factors from a slightly different approach than what one might anticipate. N and K are chosen simply as (a) the number of total decisions modeled and (b) the number of decisions which depend upon other decisions. As they explain (p. 16):

> The model has two basic parameters, N, the total number of policy choices and K ($<$ N), the number of policy choices that each choice depends upon. More specifically, each of the choices is assumed to be binary, and choice-by-choice contributions to fitness levels are drawn randomly from a uniform distribution over [0,1] for each of the $2K+1$ distinct payoff-relevant combinations a choice can be part of. Total fitness is just the average of these N choice-by-choice fitness levels. Note that with K equal to its minimum value of 0, the fitness landscape is smooth and single-peaked: changes in the setting of one choice variable do not affect the fitness contributions of the remaining N-1 choice variables. At the other extreme, with K equal to N-1, a change in a single attribute of the organism or organization changes the fitness contribution of all its attributes, resulting in many local peaks rather than just one, with each peak associated with a set of policy choices that have some internal consistency. No local peak can be improved on by perturbing a single policy choice, but local peaks may vary considerably in their fitness levels.

However, a pure NK approach suffers from the disadvantage that all choices are assumed to be equal. To avoid this problem and to model a richer decisional landscape they employ an adjacency matrix, moving us into the familiar Carter Butts (1997, 2000, 2001, 2003a, 2003b) territory of connected graphs (Carley and Butts, 1997; Butts, 2000), and formal, axiomatically determined complex systems.[1]

## Adjacency Matrices

With respect to this process, in Kauffman's original NK-Boolean dynamic fitness landscape model all of the potential choices that a firm could make were considered to be equally important and the search for higher levels of fitness was carried out through a random walk across the fitness landscape (Meyer, 1996). However, under these conditions the NK model could not account for the asymmetric relationship between strategic choices that decision-makers faced. In order to better represent the asymmetric nature of the choices facing a corporation (or, as in our case, a terrorist organization), Ghemawat and Levinthal replace the interactive parameter K, as described above, with an adjacency matrix:

> How different choices (the vertices in the graph) are linked (the lines in the graph). In such a matrix, choice variable j's effect on other variables is represented by the patters of 0s and 1s in column j, with a value of 1 indicating that the payoff to the variable in the row being considered is contingent on variable j, and a value of 0 denoting independence. Similarly, reading across row i in such a matrix indicates the variables the payoff of choice variable i is itself contingent upon. The principle diagonal of an adjacency matrix always consists of 1's, but the matrix itself need not be symmetric around that diagonal.

---

[1] Replacement of the interactivity parameter, K, with an adjacency matrix is meant to let us generalize the NK approach in the directions presently of interest. A few general observations can be made about special types of graphs and the fitness landscapes that they induce over the choices and linkages they embody. Thus, given disconnected graphs, fitness landscapes are smooth as the choices corresponding to disconnected vertices are varied—irrespective of the values of other variables. Such vertices therefore lend themselves to the notion of universal (and uncontingent) best practices. And for star graphs, in which one central choice influences the payoffs from each of N-1 peripheral choices but other linkages among choices are absent (corresponding to an adjacency matrix with 1's in the first column and along the principal diagonal and 0s everywhere else), getting the first choice right is sufficient, in conjunction with a standard process of local search in an invariant environment, to lead the organization to the global optimum. But what about graphs more generally? Exhaustive enumeration of all the graphs with N vertices and analysis of their fitness landscapes is unlikely to prove productive for even moderately large N: the number of 6-vertice graphs is 157, 7-vertice graphs 1,044, 8-vertice graphs 12,346, and so on. Restricting attention to connected graphs doesn't help much with the numbers problem since the number of connected graphs grows much more quickly than the number of disconnected graphs: with N equal to 5, disconnected graphs account for about 38% of the total, but with N equal to 8, that figure is down to less than 10%. We therefore pass up the opportunity to engage in exhaustive (and exhausting) enumeration. We begin, instead, by considering two classes of adjacency matrices that highlight two fundamental sources of asymmetry among choices, in terms of hierarchy and centrality, and comparing them with the canonical NK structure on which previous work has focused. (pp. 17-18).

In order to simplify their examination of the relationship between asymmetric choices Ghemawat and Levinthal elected to look at two adjacency matrices that highlight the classical types of choice asymmetry: hierarchy, and centrality. In the hierarchical matrix choice 1 is the most important influencing all other choices below it choice two is the second most important and so on to the final choice (in this case choice 10 were N=10) which is influences by all proceeding choices but influences only itself. For the centrality matrix choice 1 is the most central both influencing and also being influenced by all other possible choices, choice two is the second most central being influenced by all other choices and influencing all choices with exception 10 and so on. These two matrices are benchmarked against a traditional NK structure. The matrix in this case is structured such that there will be K 1's in each row and column but they will be randomly distributed across the matrix.[2] In their simulation, Ghemawat and Levinthal put K=6 which proved the same number of peaks at the other two matrices [Ghemawat and Levinthal 2005]

## Hierarchy

Ghemawat and Levinthal treat hierarchical decisions as directed trees where the 1 appears to the left of the principal diagonal. In this regard as we have explained above, choice 1 is the most hierarchically important, choice 2 the second most important, etc. (Figure 4a, Ghemawat and Levinthal, 2000).

$$
\begin{array}{c}
\phantom{00}1\ 2\ 3\ 4\ 5\ 6\ 7\ 8\ 9\ 10 \\
\begin{array}{r}
1 \\ 2 \\ 3 \\ 4 \\ 5 \\ 6 \\ 7 \\ 8 \\ 9 \\ 10
\end{array}
\left(
\begin{array}{cccccccccc}
1 & 0 & 0 & 0 & 0 & 0 & 0 & 0 & 0 & 0 \\
1 & 1 & 0 & 0 & 0 & 0 & 0 & 0 & 0 & 0 \\
1 & 1 & 1 & 0 & 0 & 0 & 0 & 0 & 0 & 0 \\
1 & 1 & 1 & 1 & 0 & 0 & 0 & 0 & 0 & 0 \\
1 & 1 & 1 & 1 & 1 & 0 & 0 & 0 & 0 & 0 \\
1 & 1 & 1 & 1 & 1 & 1 & 0 & 0 & 0 & 0 \\
1 & 1 & 1 & 1 & 1 & 1 & 1 & 0 & 0 & 0 \\
1 & 1 & 1 & 1 & 1 & 1 & 1 & 1 & 0 & 0 \\
1 & 1 & 1 & 1 & 1 & 1 & 1 & 1 & 1 & 0 \\
1 & 1 & 1 & 1 & 1 & 1 & 1 & 1 & 1 & 1
\end{array}
\right)
\end{array}
$$

---

[2] In most cases Ghemawat and Levinthal use a Poisson distribution or another, uniform distribution, noting, in any case that the probability distribution is not likely to be a mathematically relevant factor in the overall distribution of decision outcomes (i.e., the probability distribution is not the determinative property).

## Centrality

In contrast, their treatment of centrality involves interconnected decisions and, hence produces an almost perfect 90 degree rotated distribution (Figure 4b, Ghemawat and Levinthal, 2000):[3]

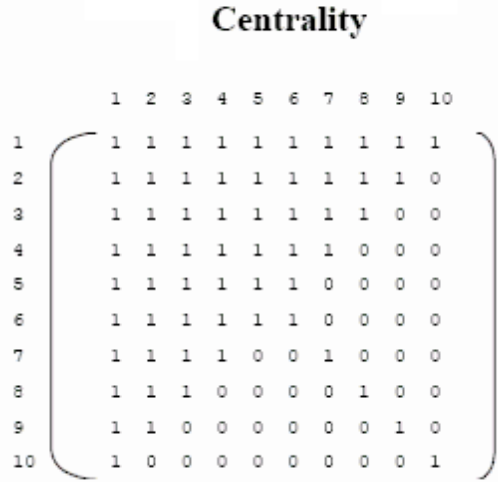

## Modeling Policy Choices

Levinthal and Ghemawat then benchmark what happens in these two types of structures against the random (but symmetric) activity which is built into the canonical NK structure. As they explain:

> For all three structures, an organization's policy choices are represented by a vector of length $N$ where each element of the vector can take on a value of 0 or 1 (not to be confused with the 0s and 1s assigned, respectively, to denoting the absence or presence of linkages between every pair of policy elements). The overall fitness landscape will then consist of $2^N$ possible policy choices, with the overall behavior of the organization characterized by a vector $\{x_1, x_2, ..., x_N\}$, where each xi takes on the value of 0 or 1. If the contribution of a given element, $xi$, of the policy vector to the overall payoff is influenced by $Ki$ other elements—in ways that vary across the three structures we will analyze—then it can be represented as $f(xj|xi1, xi2, . . . , xiKi)$. Therefore, each element's payoff contribution can take on $2Ki+1$ different values, depending on the value of the attribute itself (either 0 or 1) the value of the $Ki$ other elements by which it is influenced (each of these $Ki$ values also taking on a value of 0 or 1) and—less commonly highlighted—the luck of the draw. Specifically, it is common to assign a random number drawn from the uniform distribution from zero to one to each possible $f(xj|xi1, xi2, . . ., xKi)$ combination with the overall fitness value then being defined as
> $\Sigma i=1$ to N $f(xi|xi1, xi2, ..., xiKi) / N$. (pp. 19-20)

Their simulation structure assumes for the random benchmark that K=6, primarily because this value generates roughly the same number of local peaks as the hierarchical and central distributions.[4] Similarly

---

[3] (p. 19) "The particular form of hierarchy we explore in this paper has 1s as all the entries to the left of the principal diagonal (see Figure 4a). Choice 1 is hierarchically the most important, choice 2 the second most important, and so on. In contrast, in a set of interaction patterns ordered by a centrality measure, policies vary in terms of their interdependence with other policy choices and this interdependence is taken to be symmetric (to distinguish it as sharply as possible from the one-way influences of hierarchy). As a result, the 1s to the left of the principal diagonal are mirrored by 1s to its right. Whether the 1s cluster centrally in the adjacency matrix, however, depends on the order in which choice variables are labeled. The particular form of centrality we explore in this paper embodies a structure and a labeling scheme that has 1s as all the entries to the left of the inferior diagonal (but distributed symmetrically to the left and the right of the principal diagonal)—see Figure 4b. Thus, choice 1 is most central, choice 2 second most central, and so on.

[4] Ghemawat and Levinthal provide five additional caveats, of which the three important for our purposes are: "A number of important assumptions, based on prior applications, are built into this specification. First of all, there is the emphasis on choice under

they set N = 10, which is sufficient to generate more than a million distinct graphs, which allows them to report results averaged over a thousand independent landscapes which share the same structure. These landscapes will be either hierarchical, central or random, characterized by the particular adjacency matrix structure for each type, but with a distinct seeding (0,1) from a uniform random distribution for the fitness of the policy variables.

## Analyzing the Results of the Ghemawat-Levinthal Simulation

The first question which the authors ask "what are the effects of presetting a certain number of policy choices equal to their values at the global optimum with the remaining choices determined by a process of local search?" is interesting from a complexity science point of view, but not immediately obvious in its application to terrorist organizations. The reason for this is that while answering this question allows Ghemawat and Levinthal to address issues of strategic planning and "grand strategy" in business organizations, it doesn't really provide a reliable guide for the $C^4I$ functioning of terrorist organizations. If this were all that their simulation achieved, it would have rather limited interest for us. However their second question "what happens when one of the N values of the policy variable is preset to a value inconsistent with the value of that variable for the global peak?" is of very substantial interest as it speaks to exactly the kind of distortions which we would wish to induce in the terrorist decision-making chain.

The first interesting result of the simulation is the difference between an optimal preset of policy configurations, the hierarchical, the central and the random simulation arrays:[5]

---

uncertainty. In addition to its arguable descriptive realism, initial uncertainty helps explain why an organization launched over a fitness landscape may not instantly alight on the globally optimal policy vector. Second, there is the assumption that randomness takes the form of a uniform distribution. While some might argue that this distribution is too diffuse, we retain this assumption to provide at least some basis for numerical comparability with prior work, which suggests, among other things, that the structure of the fitness landscape is not sensitive to the particular probability distribution employed (Weinberger, 1991). Third, there is the equal weighting of different choices in terms of their direct contribution (potential) to overall fitness. Again, we retain this prior assumption even though we intend to focus on asymmetries among choices. Putting different weights on the direct contributions of choice elements does not seem to us to be the best way of gaining insight into the indirect contributions that choice elements can make to overall performance by virtue of the linkages among them. (p. 21)

[5] Ibid.

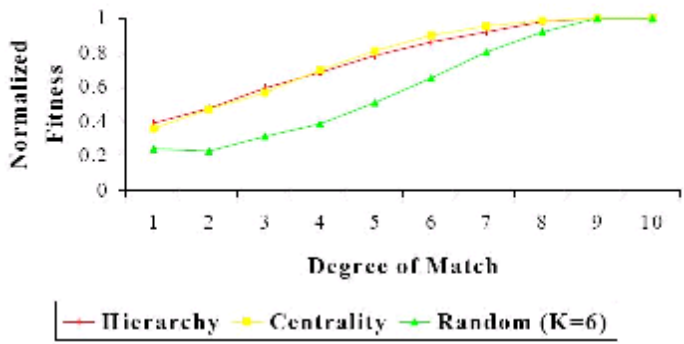

Figure 5. Value of Partially Articulated Activity Maps

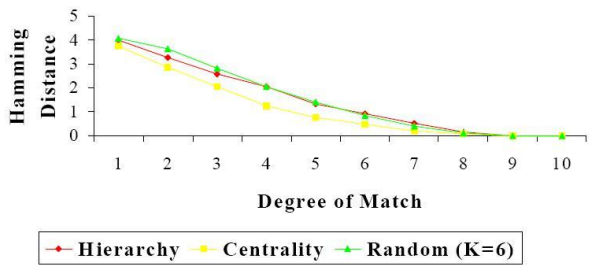

Figure 6. Partially Specified Activity Maps and Proximity to Global Optimum

Ghemawat and Levinthal describe this as:

> With a degree of match of 1, only the first, most strategic, variable is set equal to the global optimum. As more variables are matched with their settings at the global optimum, fitness rises steadily according to **Figure 5.** However, the global optimum is not approached until nearly all policy variables are specified to equal their settings at the global optimum. Similarly, hamming distances tend to be quite large (see **Figure 6**).
> The gap between the curve depicting performance under the random network structure and the other two curves indicates the power of presetting more strategic variables to their values at the global optimum. In contrast, the gap between the realized fitness level and the value of 1, indicates the loss from the not fully articulating the optimal policy array.

Admittedly, while this is an interesting result, so far it does not tell us much more than the intuitively plausible idea that an inability or failure to articulate global optima means that you probably never get there. There is, however, a slightly more interesting subtext here with respect to the complexity of rugged fitness landscapes, something originally articulated by Farmer, Packard and Kauffman ("The Structure of Rugged Fitness Landscapes" in Kauffman, 1993):

> To make more sense of these patterns, it is useful to note that the fitness landscapes we are analyzing are quite complex, typically comprising over 40 local peaks. In such worlds, the powers of local search are relatively limited. Local search rapidly leads to the identification of a local peak but conveys no assurance about the local peak's global properties (i.e., its fitness value relative to the global optimum). Presetting the most strategic variables to their values at the global optimum does lead to the identification of a better-than-average local peak (recall that the normalized fitness value would have a value of zero if the average realized fitness level equaled the average value of local peaks in the fitness landscape). However, a high level of specificity is necessary to obtain the highest possible fitness levels or configurations close to the global optimum: in rugged landscapes, there are just too many positive-gradient paths that lead to local peaks other than the global one. (p. 26)

What is perhaps most interesting here are the suggestions that (1) the events of 9-11 are likely to prove the exception, rather than the rule (also giving some optimism to the long-run possibility of ultimately negating terrorist threats, something as unthinkable today as the end of the Cold War was forty or fifty years ago) and (2) the indication that even very hierarchical terrorist organizations are unlikely to reach an optimal policy set through local search (and, following Kauffman, there simply are no other search mechanics available). Figure 7 provides what may be a slightly more interesting insight. Much in the fashion of Lissack (1996), Ghemawat and Levinthal use an approach similar to "patching" in order to simulate a feedback situation where initial errors in policy choice can be corrected and the corrections incorporated into subsequent searches. They note that (p. 26):

> …while the articulation of and insistence on adherence to a single (or low-dimensional) strategic choice may not be sufficient to lead to the identification of a high-performing set of choices, a lack of such strategic discipline is likely to lead to even less attractive results. Compare the top line in **Figure 7,** tracing the value of partially articulated activity maps in a hierarchical context in which preset choices cannot be varied (á la **Figure 5**) with the bottom line, which looks at a hierarchical context in which the preset policy choices *can* be revised in the process of local search. It turns out that with the degree of match of 1, the latter, "unconstrained" approach underperforms the "constrained" approach, and the gap between the two widens for intermediate degrees of match prior to convergence as the degree of match hits 10. Similarly, the unconstrained approach fails to generate smaller hamming distances than the constrained approach. In that sense, strategic discipline *is* useful.

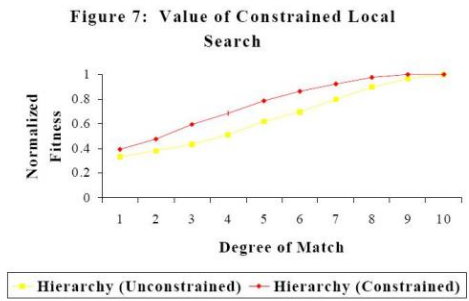

In this regard, as we have previously suggested (Fellman, Wright and Sawyer, 2003; Fellman and Wright, 2003) forcing compartmentation of terrorist cells and interrupting communications between command echelons is likely to significantly impede the ability of those cells to function effectively. From a counter-terrorism perspective, however, the most interesting part of the simulation is the section which deals with "the constraints of history". In this case, Ghemawat and Levinthal artificially inject a variable inconsistent with the global maximum and then observe how the three different types of organization adapt to this distortion. Their Figure 8 summarizes the (normalized) fitness levels achieved in the simulation when they preset a variable to a value which is inconsistent with the global optimum this is shown below (p. 27):

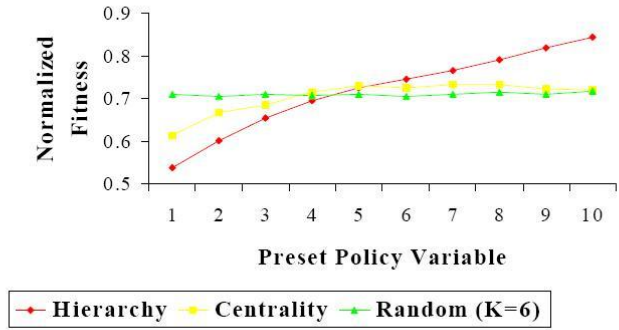

**Figure 8. Constraints of History**

The green line, representing a random pattern of interactions, not surprisingly has a fitness which is essentially "history independent". The more surprising finding is that while under conditions of hierarchy the fitness level changes quite rapidly as the mismatched variable is shifted from the most important decision to decisions of less importance, centrality conveys virtually no "recovery advantage" at all. This is a striking contrast to Porter (1996) who argues that under conditions of "third order strategic fit" (strong interlinkage/strong centrality) the decision chain becomes "as strong as its strongest link". In dealing with terrorist organizations, which are primarily hierarchical in nature (Sageman, 2004), what this finding says is that disinformation is a useful tactic (or strategy) only if it succeeds in influencing one of the key decisional variables. In other words, disinformation at the local level is unlikely to have any lasting impact on terrorist organizations. This finding also challenges the institutional wisdom of assigning case officers in the field to this type of counter-terrorism operation (Gerecht, 2001; Codevilla, 2004b).[6] In fact, from an operational point of view, the hierarchical nature of terrorist organizations means that there may be something of a mismatch in the entire targeting process. As Ghemawat and Levinthal note, "Less central variables not only do not constrain, or substantially influence the payoff of many other choices, but they themselves are not greatly contingent upon other policy choices. Being contingent on other policy choices facilitates compensatory shifts in policy variables other than the one that is preset. As a result of the absence of such contingencies, the presetting of lower-order policy choices is more damaging to fitness levels under the centrality structure." (p. 27) The problem, however, is that decisionmaking in terrorist organizations is apt to be operating under conditions of hierarchy rather then centrality. Perhaps even more annoyingly, those organizations operating against terrorist cells may themselves be organized in a more modern fashion, availing themselves of divisional interlinkages and flat management structure, so that the mismatch between operational objectives and terrorist organizations may, in fact, prove organizationally damaging to the counter-terrorist organization both in a relative and an absolute sense. In this regard, pointless, or fruitless counter-terrorism operations, particularly when conducted at the field level may do considerably more damage than good, particularly to the organization striving to combat terrorism. This is a feature of counter-terrorism that is not entirely unfamiliar to case officers who have operated in this capacity (Gilligan, 2003; Gerecht 2001).

---

[6] Ghemawat and Levinthal test this another way and come to essentially the same conclusion: "Another striking feature of this set of simulations concerns how few of the optima with preset mismatches constitute local peaks of the fitness landscape. Given the importance of configurational effects, one might reasonably conjecture that constraining one variable to differ from the global optimum would lead to the selection of a different, non-global, peak in the fitness landscape. However, Figure 9 indicates that this is relatively uncommon except as one turns to presetting the least important variables under the hierarchy and centrality structures." (pp. 28-29)

In their final section, Ghemawat and Levinthal examine choice structures.[7] To do this, they set up three classes of variables: independent variables (1-3), whose payoff is not dependent upon that of any other variables, but which influences the payoff of the dependent variables (4-6), and three variables (7-9) which are simply independent of all others (Figure 12).

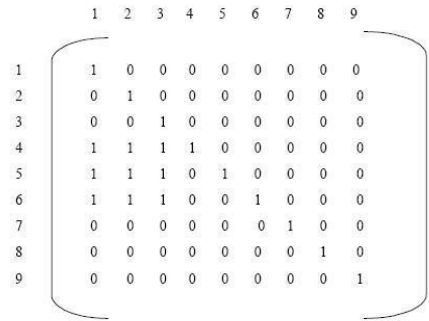

They illustrate their findings in Figure 13:

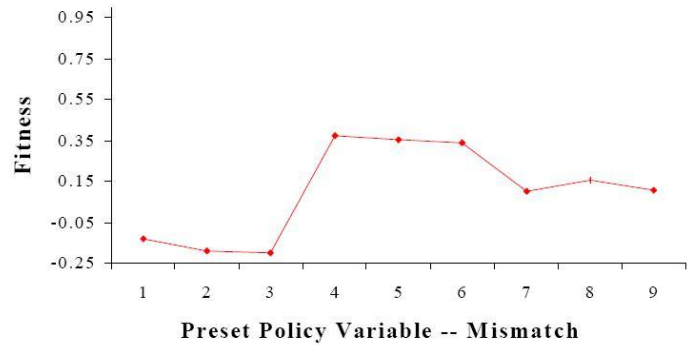

Interestingly, Ghemawat and Levinthal are surprised that constraining the independent variables to values different from the global maximum affects overall fitness more than the constraint of contingent variables:

> **Figure 13** indicates that constraining one of the "influential" variables to differ from the global maximum has a profound effect on the relative fitness level of the constrained optimum. Surprisingly, constraining the independent variables to differ from the global optimum has a larger impact than constraining the seemingly more important "contingent" variables. The reason for this is that the presence of contingency allows for the possibility of substituting or compensating changes in policy variables. While tightly linked interaction patterns have generally been viewed as fragile, the equifinality that high levels of interaction engender also allows for a certain robustness. In contrast, when an autonomous variable is misspecified, that doesn't create negative

---

[7] "The broader suggestion is that the 'natural' adjacency matrices we have looked at so far mix up at least three very different types of effects: influence, contingency and autonomy. Variables may be more or less influential to the extent that they affect the payoffs to other variables. In an adjacency matrix, this is represented by the prevalence of 1s in the relevant column. Independent of influence, the payoffs from specific variables may be more or less contingent on other choices, as reflected in the number of 1s in the relevant row of the adjacency matrix. And autonomy is characterized by variables that are neither influential nor contingent: variables that correspond, in graph-theoretic terms, to disconnected vertices. In this subsection, we look at a choice structure—distinct from the three that we have already examined—that distinguishes particularly clearly among these three effects." (p. 29)

> ramifications elsewhere in the system of policy choices; at the same time, however, there is no opportunity to compensate for the misspecification.

The most probable source of their surprise is the literature on strategic linkages (which they cite heavily at the beginning of their study).

## Conclusion

What the Ghemawat-Levinthal simulation ultimately reveals is that coupled decisional structures have a compensatory feature which makes them less vulnerable to certain kinds of decisional disruption than hierarchical decision structures. The implicit lesson for intelligence organizations is that in dealing with terrorist organizations efforts would best be directed at hitting them hard, at the highest levels of the hierarchy and leave the "cleanup" to local law enforcement organizations. In this context, less engagement may, in fact, yield more results. Conversely, if "sometimes less is more", clearly "sometimes more is a whole lot less."

# References

Butts, Carter T. (2000). "An Axiomatic Approach to Network Complexity." Journal of Mathematical Sociology, 24(4), 273-301.

Butts, Carter T. (2001). "The Complexity of Social Networks: Theoretical and Empirical Findings." Social Networks, 23(1), 31-71

Butts, Carter T. (2003a) "Network Inference, Error, and Informant (In)Accuracy: A Bayesian Approach", Social Networks 25(2) 103-30.

Butts, Carter T. (2003b). "Predictability of Large-scale Spatially Embedded Networks." In Ronald Breiger, Kathleen Carley, and Philippa Pattison (eds.), Dynamic Social Network Modeling and Analysis: Workshop Summary and Papers, 313-323. Washington, D.C.: National Academies Press.

Carley, Kathleen M. and Butts, Carter T. (1997). "An Algorithmic Approach to the Comparison of Partially Labeled Graphs." In Proceedings of the 1997 International Symposium on Command and Control Research and Technology. June. Washington, D.C.

Carley, Kathleen; Lee, Ju-Sung; and Krackhardt, David; (2001) "Destabilizing Networks", Connections 24(3): 31-34, INSNA (2001)

Carley, Kathleen (2002a)  Modeling "Covert Networks", Paper prepared for the National Academy of Science Workshop on Terrorism, December 11, 2002

Carley, Kathleen (2002b) "Inhibiting Adaptation", Proceedings of the 2002 Command and Control Research and Technology Symposium, Naval Postgraduate School, Monterey, CA (2002)

Carley, Kathleen; Dombroski, Matthew; Tsvetovat, Maksim; Reminga, Jeffrey; Kamneva, Natasha (2003) "Destabilizing Dynamic Covert Networks", Proceedings of the 8$^{th}$ International Command and Control Research and Technology Symposium, National Defense War College, Washington, D.C. (2003) http://www.casos.cs.cmu.edu/publications/resources_others/a2c2_carley_2003_destabilizing.pdf

Carley, Kathleen; Diesner, Jana; Reminga, Jeffrey; and Tsvetovat,, Maksim (2004), "Toward an end-to-end approach for extracting, analyzing and visualizing network data", ISRI, Carnegie Mellon University (2004)

Clemens, Jonathan P. and O' Neill, Lauren "Discovering an Optimum Covert Network", Santa Fe Institute, Summer, 2004

Codevilla, Angelo (2004a) "Doing it the Hard Way", Claremont Review of Books, Fall, 2004 http://www.claremont.org/writings/crb/fall2004/codevilla.html

Codevilla, Angelo (2004b) "Why U.S. Intelligence is Inadequate and How to Fix It", Center for Security Policy, Occasional Papers Series, December, 2004 http://www.centerforsecuritypolicy.org/occasionalpapers/Why-US-Intelligence-Is-Inadequate.pdf

Fellman, Philip V.; Sawyer, David; and Wright, Roxana (2003) "Modeling Terrorist Networks - Complex Systems and First Principles of Counter-Intelligence," Proceedings of the NATO Conference on Central Asia: Enlargement, Civil – Military Relations, and Security, Kazach American University/North Atlantic Treaty Organization (NATO)  May 14-16, 2003

Fellman, Philip V. and Wright, Roxana (2003) "Modeling Terrorist Networks: Complex Systems at the Mid-Range", paper prepared for the Joint Complexity Conference, London School of Economics, September 16-18, 2003  http://www.psych.lse.ac.uk/complexity/Conference/FellmanWright.pdf

Fellman, Philip V. and Strathern, Mark (2004) "The Symmetries and Redundancies of Terror: Patterns in the Dark", Proceedings of the annual meeting of the North American Association for Computation in the Social and Organizational Sciences, Carnegie Mellon University, June 27-29, 2004. http://casos.isri.cmu.edu/events/conferences/2004/2004_proceedings/V.Fellman,Phill.doc

Gerecht, R,.M., "The Counterterrorist Myth", The Atlantic Monthly, July-August, 2001

Ghemawat, Pankaj and Levinthal, Daniel (2000) "Choice Structures, Business Strategy and Performance: A Generalized NK-Simulation Approach", Reginald H. Jones Center, The Wharton School, University of Pennsylvania (2000) http://www.people.hbs.edu/pghemawat/pubs/ChoiceStructures.pdf

Gilligan, Tom CIA Life: 10,000 Days with the Agency, Intelligence E-Publishing Company (November 2003)

Hoffman, Bruce (1997) "The Modern Terrorist Mindset: Tactics, Targets and Technologies", Centre for the Study of Terrorism and Political Violence, St. Andrews University, Scotland, October, 1997

Hoffman, Bruce and Carr, Caleb (1997) "Terrorism: Who Is Fighting Whom?" World Policy Journal, Vol. 14, No. 1, Spring 1997

Kauffman, Stuart (1993), The Origins of Order, Oxford University Press (Oxford and New York: 1993)

Kauffman, Stuart (1996), At Home In the Universe, Oxford University Press (Oxford and New York: 1996)

Kauffman, Stuart (2000) Investigations, Oxford University Press (Oxford and New York, 2000)

Krebs, Valdis (2001) "Uncloaking Terrorist Networks", First Monday, (2001) http://www.orgnet.com/hijackers.html

Lissack, Michael, (1996)  "Chaos and Complexity: What Does That Have to Do with Knowledge Management?", in Knowledge Management: Organization, Competence and Methodology, ed. J. F. Schreinemakers, Ergon Verlog 1: 62-81 (Wurzburg: 1996)

McKelvey, Bill (1999) "Avoiding Complexity Catastrophe in Coevolutionary Pockets: Strategies for Rugged Landscapes", Organization Science, Vol. 10, No. 3, May–June 1999 pp. 294–321

Meyer, Chris "What's Under the Hood: A Layman's Guide to the Real Science", Center for Business Innovation, Cap Gemini, Ernst and Young, Conference on Embracing Complexity, San Francisco, July 17-19, 1996

Porter, Michael (1996) "What is Strategy?", Harvard Business Review, November-December, 1996

Sageman, Mark (2004) Understanding Terror Networks, University of Pennsylvania Press, 2004.

Tsvetovat, Max & Carley, Kathleen. (2003). Bouncing Back: Recovery mechanisms of covert networks. NAACSOS Conference 2003